# Significant *ZT* Enhancement in p-type Ti(Co,Fe)Sb-InSb Nanocomposites via a Synergistic '*High Mobility Electron Injection, Energy filtering and Boundary Scattering*' Approach

Wenjie Xie[a, b], Yonggao Yan[a], Song Zhu[c], Menghan Zhou[c], Sascha Populoh[b], Krzysztof Gałązka[b], S. Joseph Poon[d], Anke Weidenkaff[b, *], Jian He[c, *], Xinfeng Tang[a,*], and Terry M. Tritt[c]

a) State Key Laboratory of Advanced Technology for Materials Synthesis and Processing, Wuhan University of Technology, Wuhan 430070, People's Republic of China

b) EMPA, Swiss Federal Laboratories for Materials Science and Technology, Solid State Chemistry and Catalysis, CH-8600 Düebendorf, Switzerland

c) Department of Physics and Astronomy, Clemson University, Clemson, SC 29634-0978, USA

d) Department of Physics, University of Virginia, Charlottesville, VA 22904-4714, USA

It has been demonstrated that InSb nanoinclusions, which are formed *in situ,* can simultaneously improve all three individual thermoelectric properties of the *n*-type half Heusler compound (Ti,Zr,Hf)(Co,Ni)Sb [Xie *et al.*, Acta Mater. 58, 4795 (2010)]. In the work presented herein, we have adopted the same approach to the *p*-type half Heusler compound Ti(Co,Fe)Sb. The results of resistivity, Seebeck coefficient, thermal conductivity, and Hall coefficient measurements indicate that the combined *high mobility electron injection, low energy electron filtering, and boundary scattering,* again, lead to a simultaneous improvement of all three individual thermoelectric properties: enhanced Seebeck coefficient and electrical conductivity as well as reduced lattice thermal conductivity. A figure of merit of *ZT* ~ 0.33 was attained at 900 K for the sample containing 1.0 at.% InSb nanoinclusions, a ~ 450% improvement over the nanoinclusion-free sample. This represents a rare case that the same nanostructuring approach successfully works for both *p*-type and *n*-type thermoelectric materials of the same class, hence pointing to a promising materials design route for higher performance half-Heusler materials in the future and hopefully will realize similar improvement in TE devices based on such half Heusler alloys.

**Keywords**: thermoelectric; half-Heusler; nanocomposites

# Introduction

Thermoelectricity is the simplest technology applicable to the direct conversion of waste heat into electricity in an environmentally friendly way. A practical thermoelectric (TE) device consists of legs each made of high dimensionless figure of merit (*ZT*) *n*-type or *p*-type material, where $ZT=\alpha^2\sigma T/\kappa$, $\alpha$ is the Seebeck coefficient, $\sigma$ the electrical conductivity, $\kappa$ the thermal conductivity, and T the absolute temperature. The thermal conductivity is comprised of two parts; one due to the charge carriers ($\kappa_e$) and one due to the lattice ($\kappa_L$); $\kappa = \kappa_e + \kappa_L$. Though there is no known upper limit for ZT theoretically, the state-of-the-art TE materials have a ZT ~ 1-2 because of the inter-dependence of $\alpha, \sigma,$ and $\kappa$. A *ZT* value ~ 1 has been regarded as the benchmark for practical TE materials. Among the wide variety of TE materials, half-Heusler (HH) compounds have attracted considerable interest because of their promising *ZT* values between 600 K and 1000 K, which happens to also be the temperature range of *most industrial waste heat sources* and thus are quite suitable for TE energy harvest in this temperature range. [1-4] HH compounds crystallize in the cubic MgAgAs type structure featured by four interpenetrating face-centered-cubic (*fcc*) sublattices, allowing for doping on each of the three occupied *fcc* sublattices in order to manipulate each of the individual TE properties. [5-9] In *n*-type MNiSn (M = Ti, Zr, and Hf) based HH compounds, for example, the carrier concentration can be tuned via doping the Sn-site while the lattice thermal conductivity can be reduced via doping the M- and/or Ni-sites, thereby doping induced point defects, mass fluctuations and strain field fluctuations in order to effectively scatter short- and mid-wavelength heat-carrying phonons. *ZT* values ~ 1.0 at 725 K, ~ 0.81 at 1025 K, and ~ 1.0 at 1000 K have been reported for *n*-type ($Ti_{0.37}Zr_{0.37}Hf_{0.26}$)NiSn [10], ($Hf_{0.75}Zr_{0.25}$)Ni($Sn_{0.975}Sb_{0.025}$), [11] and ($Hf_{0.6}Zr_{0.4}$)Ni($Sn_{0.98}Sb_{0.02}$) [12], respectively.

A practical HH compound based TE device requires not only high *ZT* *n*-type and *p*-type materials but also that the *n*-type and *p*-type materials have similar composition and thus similar thermal expansion and mechanical properties. However, the *ZT* values of *p*-type HH compounds have previously been inferior to those of *n*-type counterparts due to the fact that doping can only optimize the TE properties to a fairly limited degree. For example, *p*-type MCoSb HH compounds can be prepared via doping the Co- and/or Sb-site, e.g., TiCo$_{0.85}$Fe$_{0.15}$Sb [13], Zr$_{0.5}$Hf$_{0.5}$CoSb$_{0.8}$Sn$_{0.2}$, [14] and Zr$_{0.5}$Hf$_{0.5}$Co$_{0.3}$Ir$_{0.7}$Sb$_{0.99}$Sn$_{0.1}$ [15]. The highest *ZT* ~ 0.5 at 973 K was reported for *p*-type Zr$_{0.5}$Hf$_{0.5}$CoSb$_{0.8}$Sn$_{0.2}$ [14].

Further improvement of the TE properties of *p*-type HH compounds is thus dependent upon new materials design approaches other than simple substitutional doping. To this end, the nanostructuring process is a promising approach: since nanostructuring has become a new paradigm approach of improving the *ZT* of bulk TE materials over the past several years. In a nanostructured bulk TE material (i.e., nanocomposite), the lattice thermal conductivity can be depressed by phonon scattering due to the numerous interfaces; whereas the Seebeck coefficient can be increased, without significantly sacrificing the electrical conductivity, by an energy filtering process on the interfaces [16-18]. Implementation of the nanostructuring process in HH compounds has indeed led to enhancement of the *ZT* in many cases, which has been predominantly due to a significant reduction of the lattice thermal conductivity [19-22]. Until recently, Xie *et al.* [23] and Makongo *et al.* [24] reported that nanostructuring process simultaneously enhanced the Seebeck coefficient and electrical conductivity, and depressed the thermal conductivity in *n*-type (Ti,Zr,Hf)(Co,Ni)Sb-InSb and Zr$_{0.25}$Hf$_{0.75}$NiSn-Zr$_{0.25}$Hf$_{0.75}$Ni$_2$Sn nanocomposites, respectively. An immediate question arises as to whether one can apply the same nanostructuring process in *p*-type HH compounds, and whether a positive outcome can be realized.

We report herein that an inductive-melting-spark-plasma-sintering synthesis process of *p-type* Ti(CoFe)Sb based nanoscomposites in which InSb nanoinclusions were formed *in situ* during the synthesis process. We have confirmed experimentally that this process results in *in* *situ* formed InSb nanoinclusions which in effect simultaneously enhanced the Seebeck coefficient and electrical conductivity while reducing the lattice thermal conductivity. We attribute these changes to the combined nanoinclusion-induced *high mobility electron injection, energy filtering, and boundary scattering*. In view of our previous work on the *n*-type HH compounds [23], the current study thus sheds new light on employing the nanostructuring process in order to enhance the electrical properties of HH compounds and other classes of TE materials.

## Experimental procedures

Ingots of TiCo$_{0.85}$Fe$_{0.15}$Sb alloys with nominal ratio x = 0, 0.7, 1, 1.5 and 3 at.% of InSb nanoinclusions that formed *in-situ*, were prepared by inductively melting appropriate amounts of single elements in an argon atmosphere. To compensate for the evaporation of Sb, a 5 at.% excess of Sb powders was added in the starting materials. All the ingots were melted three times, and then annealed at 1173 K for a week to ensure sample homogeneity. The ingots were then pulverized and sintered using a spark plasma sintering (SPS) technique at 1373 K for 8 min under the pressure of 45 MPa in order to obtain highly densified bulk samples.

The density (*d*) of SPSed bulk samples was determined by the Archimedes' method and found to be at least 97% of the theoretical density. The phase structure, chemical composition and micro-morphology of the bulk samples were analyzed by means of X-ray diffraction (XRD) (PANalytical X'Pert Pro ® X-ray diffractometer), electron probe microanalysis (JXA-8230®), and field emission scanning electron microscopy (SEM, Hitachi S-4800®) with Energy-dispersive X-ray spectroscopy (EDS). The bulk samples were cut into rectangular bars with dimensions ~ 2 × 2 × 10 mm$^3$ for electrical conductivity and Seebeck coefficient measurements on a ZEM-3 system (Ulvac Riko Inc.) under an inert gas (He) atmosphere from 300 to 900 K. The thermal diffusivity was measured by the laser flash method on a Netzsch LFA457® system. Specific heat was measured by differential scanning calorimetry on a Netzsch DSC-404C® calorimeter. The thermal conductivity, $\kappa$, was then calculated from the thermal diffusivity, $D$, specific heat per unit volume, $C_p$, and density, $d$, from the relationship, $\kappa = DC_p d$. The electrical conductivity, Seebeck coefficient and thermal conductivity measurements on the same samples were also conducted on our custom-designed low-temperature apparatuses from 15 to 310 K [25–26]. The low temperature specific heat data was taken on a Quantum Design® Quantum Design Physical Properties Measurement System (PPMS) from 1.8 K to 300 K. Hall coefficients, $R_H$, were measured on a PPMS using a five-probe configuration by sweeping the magnetic field between ± 1.5 T from 5 K to 400 K. The effective carrier concentration, $n$, and Hall mobility, $\mu$, were then calculated from the relations $n = 1/R_H e$ and $\mu = \sigma/ne$, respectively, where $e$ is the elemental charge. Uncertainties in the low temperature electrical conductivity, Seebeck coefficient and thermal conductivity measurements are ≈ ±5%, ±2% and ±7%, respectively, and primarily due to the accurate determination of sample dimensions. However, by making these low temperature measurements on the same sample the cross sectional area cancels in the calculation of the $ZT$, thus greatly decreasing the uncertainty in $ZT$ to ≈ ±7%, then the high temperature data taken on the different systems can be matched to the low temperature data to reduce the overall uncertainty in $ZT$. It is worth mentioning that all the low temperature and high temperature measurements of electrical conductivity, Seebeck coefficient, specific heat, and thermal conductivity match very well.

## Results and Discussion

The powder X-ray diffraction (XRD) patterns for TiCo$_{0.85}$Fe$_{0.15}$Sb-x% InSb (x = 0, 0.7, 1.0, 1.5, and 3.0; hereafter named sample HH-x, for simplicity) composites are presented in **Fig. 1**. The main peaks can be indexed to the HH structure, and the InSb peaks are progressively enhanced with increasing x value. In addition, small amounts of FeSb$_2$ and FeSb are detected for all samples, including sample HH-0. The formation of FeSb$_2$ and FeSb phases are likely due to reaction of Fe with excess Sb. Nonetheless, as the content and micro-morphology of $FeSb_2$ and FeSb are found to be practically the same for all samples, irrespective of the content of InSb inclusions, it is thus plausible to treat $FeSb_2$ and FeSb as part of the HH matrix, if not otherwise noted.

SEM was employed to check the micro-morphology of the InSb nanoinclusions and of the coarse grained HH matrix for all samples. For simplicity, we only present and discuss the results of two representative samples HH-0 and HH-3. The typical grain size of HH matrix is ~ 30-50 μm for sample HH-0 (**Fig. 2a**), and the grain boundary is very clearly defined (**Fig. 2b**). In contrast, the average grain size of HH matrix is dramatically decreased to 5-10 μm for sample HH-3 upon the addition of Indium (**Fig 2c**). We have observed a clear correlation between the content of the InSb precipitates and average grain size of the HH matrix for all samples we investigated. Importantly, a large number of InSb precipitates (the composition was determined by EDS analysis) with a typical size of 20-60 nm are found on the boundary of HH-matrix grains (**Fig. 2d**), which is very reminiscent of similar observations in the *n*-type HH nanocomposites **[23]**. Applying the Scherrer's equation to the XRD data yields a well consistent average grain size of InSb precipitates of ≈ 60 nm for sample HH-3. The amount of InSb progressively increases as more Indium is added in the starting materials, allowing for a systematic study on the impact of InSb nanoinclusions on the TE properties of nanocomposites.

The temperature dependence of electrical conductivity of the HH-x nanocomposites is presented in **Fig. 3a**. All samples exhibit a semiconductor-like behavior. Note that the presence of InSb nanoinclusions effectively increases the electrical conductivity despite an increased number of grain boundaries: for example, the electrical conductivity of HH-0.7 reaches a value of $\sigma \approx 1.8 \times 10^4$ Sm$^{-1}$ at 900 K, almost two times higher than that of HH-0 (~ $8 \times 10^3$ Sm$^{-1}$). The results of Hall coefficient measurements indicate that the increase of $\sigma$ with InSb nanoinclusions is due primarily to the enhancement of the carrier mobility; furthermore, the value of carrier mobility tends to exhibit a trend that is inversely correlated to the carrier concentration (**Fig. 3b**). InSb is well known for its exceptionally high mobility on the order of $10^4$ cm$^2$V$^{-1}$s$^{-1}$ **[27]**, in contrast to the low mobility of *p*-type Ti(FeCo)Sb HH compounds (~ 0.5 cm$^2$V$^{-1}$s$^{-1}$ from **ref.13**). Meanwhile, the InSb nanoinclusons are well isolated (the volume ratio of InSb nanoinclusions is well below the percolation limit as evidenced in our extensive electron microscopy study) while the HH coarse grains form a conduction path. High mobility carriers from InSb are injected into low-mobility HH matrix, hence enhancing the overall carrier mobility. Another noteworthy feature observed in **Fig. 3b** is that the carrier concentration reaches a minimum at x =0.01 and 0.015, whereas the mobility reaches the highest values, and also where the *ZT*, as will be shown later, reaches the maximum value.

The temperature dependence of the Seebeck coefficients of the HH-x nanocomposites is presented in **Fig. 3c**. The positive sign of the Seebeck coefficients indicates a dominant *p*-type conduction in these samples. One interesting feature is the small but broad hump below 200 K, which we believe results from a Kondo effect **[28]** rather than a phonon drag effect. The presence of InSb inclusions also enhances the magnitude of the Seebeck coefficient. In the limit of a nondegerenate electron gas, the Seebeck coefficient in a single parabolic band (SPB) is related to the carrier density *n* and scattering parameter $\lambda$ via

$$\alpha = \frac{k}{e}(2+\lambda-\eta) = \frac{k}{e}\left\{2+\lambda+\ln\left[\frac{2(2\pi m^* kT/h^2)^{3/2}}{n}\right]\right\} \qquad (1)$$

Where *k* is Boltzmann's constant, *e* the elementary charge, $\eta$ the reduced Fermi energy, $\lambda$ the energy-dependent scattering parameter, *m\** the carrier effective mass, *h* the Planck constant, and *n* the carrier concentration. **[29, 30]** The measured room temperature Seebeck coefficient versus room temperature carrier concentration is presented in **Fig. 3d** with the solid line representing the expected $\alpha \sim \ln(n^{-1})$ dependence. Although the applied model may be over-simplified, it is apparent that Seebeck coefficient of HH-x samples with InSb nanoinclusions do not follow the general trend of $\alpha \sim \ln(n^{-1})$ dependence. Such deviation suggests that the variation of the Seebeck coefficient with InSb nanoinclusions is most likely due in part to the variation of the scattering parameter. Further, the magnitude of the deviation is not proportional to the x-value, suggesting the atomic weight percentage of InSb nanoinclusion may not be the sole governing factor of the electrical transport properties. Nonetheless, InSb nanoinclusions distributed in the HH matrix grain boundaries appear to act as a potential barrier to filter out the charge carriers with lower energies in the content of energy filtering effect [16-18]; meanwhile, the increased number of grain boundaries will also scatter phonons and thus lower the lattice thermal conductivity.

The power factor (*PF*) is primarily a gauge of the electrical properties of a TE material. We thus calculate the PF from the relation $PF=\alpha^2\sigma$ and shown in **Fig. 4**. Compared to the nanoinclusion-free sample HH-0, all samples with InSb nanoinclusions exhibit a significant enhancement of the power factor. The highest power factor is $\sim 1.4\times10^{-3}$ Wm$^{-1}$K$^{-2}$ at 900 K for HH-1, a ~ 360% increase as compared with that of HH-0. This result is important in that, to date, most advances in nanocomposites are due to a significantly reduced lattice thermal conductivity rather than the improvement of the PF. As there is a progressively lesser degree for the further reduction of the lattice thermal conductivity, novel mechanisms that can increase the PF are very highly desirable.

At this point we turn to the results of the impact of InSb nanoinclusions on the thermal transport properties of these materials. The temperature dependence of the thermal conductivity of these materials is shown in **Fig 5a**. As expected, the thermal conductivity of nanocomposites is lower than that of HH-0, but the reduction is much larger in the low temperatures range between 10 K and 400 K. This is understandable in that the nanoinclusions, with the typical size of few tens of nm, will more effectively scatter mid- to long wavelength heat-carrying phonons that predominate at low temperatures. Furthermore, the thermal conductivity can be separated into the lattice thermal conductivity, $\kappa_L$, and the carrier thermal conductivity, $\kappa_e$. $\kappa_e$ can be estimated via the Wiedemann-Franz relation $\kappa_e = L_0 \sigma T$, where $L_0$ is the Lorentz number ($L_0 = 2.48 \times 10^{-8}$ W$\Omega$K$^{-2}$) for a metallic like material. In general, the value of $L_0$ depends on the values of $\eta$, $\lambda$ and T. We herein adopt a $L_0 = L = 2.0 \times 10^{-8}$ W$\Omega$K$^{-2}$, a value appropriate for degenerate semiconductors. Because the electrical conductivity values are fairly low for all samples ($\sigma < 2.4 \times 10^4$ Sm$^{-1}$), the estimated carrier thermal conductivity is no more than 10% of the total thermal conductivity, therefore in other words, $\kappa \approx \kappa_L$.

As shown in **Fig. 5b**, the $\kappa_L$ first increases and then decreases with increasing of temperature for all samples. In the context of the classical kinetic theory,

$$\kappa_L = 1/3 C_V v_s l_{ph} \tag{2}$$

where $C_V$ is the heat capacity per unit volume, $v_s$ the velocity of sound, and $l_{ph}$ the mean free path of the heat-carrying phonons. At low temperatures the $\kappa_L$ will be governed by $C_v$ and will vary roughly as $T^3$. As the temperature increases, Umklapp phonon-phonon processes become more prominent and reduce the phonon mean free path, $l_{ph}$. The $\kappa_L$ thus reach a maximum and then begin to fall exponentially with temperature as $e^{\theta_D/T}$, where $\theta_D$ is Debye temperature. At higher temperatures, the exponential fall crosses over to a $T^{-1}$ behavior, where the Umklapp phonon-phonon scattering becomes the dominant scattering mechanism. So the nanoinclusions induced reduction of $\kappa_L$ is most dominant at low temperatures and gradually loses its effectiveness at elevated temperatures as Umklapp phonon-phonon scattering dominants. The lowest $\kappa_L$ of HH-x with 1.5% InSb nanoinclusions achieves a value of $\kappa_L \approx$ 3.6 Wm$^{-1}$K$^{-1}$ at 900 K, which decreases by 20% compared with the lowest $\kappa_L$ of HH-0.

The heat capacity ($C_P$) of HH-0 and HH-0.7 are plotted in **Fig. 6**. $C_P$ values of other HH-x with InSb nanoinclusions are almost identical to that of HH-0.7. The room temperature heat capacity is found to be fairly close to the Dulong-Petit limit (~ 0.327 Jg$^{-1}$K$^{-1}$, which does not take into account the InSb nanoinclusions). We fit the $C_p$ data at low temperature range (2-5 K) by the Debye model, and the Debye temperature $\theta_D$ of 384 and 367 K are obtained for HH-0 and HH-0.7, respectively. From the Debye model, the average velocity of sound $v_a$ is derived by

$$v_a = \frac{\omega_D}{K_D} = \left(\frac{2\pi\theta_D k_B}{h}\right) \bigg/ \left(6\pi^2 n_a\right)^{1/3} \qquad (3)$$

where $n_a$ is the number of atoms per unit volume. For HH-0 and HH-0.7, we have $v_a$ = 5258 m/s and 5025 m/s, which are consistent with the literature data **[31]**. Furthermore, we estimate the phonon mean free path for HH-x by assuming: 1) $v_s$ is weakly temperature dependent; and 2) $C_P$ is roughly equal to $C_V$. Using calculated $v_s$ and measured $C_P$, the phonon mean free path $l_{ph}$ of HH-x can be obtained by equation (2), and the results are presented in **Fig. 7a**. Below 100 K, the $l_{ph}$ decreases quickly with increasing temperature. It is worth pointing out that the phonon-mean-free-path $l_{ph}$ is in the range of 10-70 nm for HH-0.7 between 10 K and 50 K (where the Umklapp processes are negligible), and are comparable to the typical sizes of the InSb nanoinclusions. At higher temperature, the $l_{ph}$ gradually decreases and approaches the scale of the lattice parameter.

In order to determine to which degree that is left for a further reduction of the lattice thermal conductivity, the minimum lattice thermal conductivity, $\kappa_{min}$, for TiCoSb compound is estimated by applying a model developed by Cahill *et al.* **[32]**,

$$\kappa_{min} = \left(\frac{\pi}{6}\right)^{1/3} k_B n_a^{2/3} \sum_i v_i \left(\frac{T}{\theta_i}\right)^2 \int_0^{\theta_i/T} \frac{x^3 e^x}{(e^x - 1)^2} dx \qquad (4)$$

where the summation is over the three polarization modes and $k_B$ the Boltzmann constant.

The cut-off frequency (in unit of Kelvin) is, $\theta_i = v_i(\hbar/k_B)(6\pi^2 n_a)^{1/3}$, where $n_a$ is the number density of atoms, $\hbar$ the reduced Planck constant, $v_i$ the sound velocity for each polarization modes. Here, the longitudinal and transverse sound velocities $v_L$ = 5691 m/s and $v_S$ = 3230 m/s are collected from literature [31]. The temperature dependence of calculated $\kappa_{min}$ for TiCoSb compound is plotted in **Fig. 8**. Because the $\kappa_L$ of HH-1.5 is the lowest one among the all the HH-x nanocomposites, it is also included in **Fig. 8** for comparison. There is clearly still some difference between the $\kappa_{min}$ of TiCoSb and the lowest $\kappa_L$ of Ti(CoFe)Sb-1.5%InSb nanocomposite at high temperature range. One question is that if the system possessed nanostructures with even smaller sizes could the lattice thermal conductivity for the Ti(CoFe)Sb-1.5%InSb nanocomposite be further reduced. However, the phonon mean free path of HH-x nanocomposite is on the order of 0.8 nm at 900 K, as shown in **Fig. 7b**, which is close to the lattice parameter of the TiCoSb compound, thus limiting any further reduction in $\kappa_L$. And, we should also note that if the size of TiCoSb matrix decreases to sizes on the order of ~1 nm the small grains will also significantly decrease the carrier mobility due to enhanced scattering of the electron [33]. As such, doping is apparently the next step to further reduce the lattice thermal conductivity at high temperatures.

Finally, the thermoelectric figure of merit *ZT* values are shown as a function of temperature in **Fig. 9a**. Because InSb nanoinclusions distributed on Ti(CoFe)Sb grain boundaries induce the *high mobility electron injection energy filtering and boundary scattering effects*, the combined effects appear to decouple the electrical and thermal transport properties of Ti(CoFe)Sb-InSb nanocomposites and simultaneously enhance the PF as well as reduce the lattice thermal conductivity $\kappa_L$, the *ZT* values of all HH-x with InSb nanoinclusions are significantly higher than that of HH-0. In particular, the highest *ZT* value of 0.33 is attained at 900 K for sample HH-1.0, which is ≈ a 450% improvement over HH-0. We summarize the strategy employed by the results shown in **Fig. 9b**, the *ZT* dependence of $\mu/\kappa_L$ clearly shows that large $\mu/\kappa_L$ will result in a higher *ZT*, and this goal is achieved primarily by introducing the InSb nanocinclusions.

Finally Before we finish, we address a general issue pertinent to nanocomposites, i.e., the thermal stability of the nanoinclusions. To this end, the electrical conductivity, Seebeck coefficient and thermal conductivity are measured for multiple times in the temperature range of 300-900 K, and the results are well reproduced.

## Conclusions

We have demonstrated the viability of preparing bulk nanostructured *p*-type Ti(Co,Fe)Sb half-Heusler composites with InSb nanoinclusions that are formed *in situ* via a high-frequency induction melting process followed by a SPS process. In view of the successful implementation of nanostrucutring process in the *n*-type half-Heusler composites that we have previously reported, this work provides an important example that the same successful materials design approach can work on both *n*- and *p*-type TE materials of the same class of materials. By controlling the *in-situ* formation of InSb nanoinclusions, we have experimentally shown that the Seebeck coefficient, electrical conductivity and thermal conductivity can be independently manipulated in a manner that significantly enhances the ZT of these materials. InSb nanoinclusions inject high mobility carriers into the matrix and thus aides in enhancing the electrical conductivity. InSb-Ti(Co,Fe)Sb interfaces act as energy barriers to filter low energy carriers, improving the Seebeck coefficient. In addition, the InSb nanoinclusions enhance the phonon scattering that results in a decrease in the lattice thermal conductivity. As a result, Ti(Co,Fe)Sb-InSb nanocomposites with 1.0% nanoinclusions (HH-1.0) achieves the highest *ZT* of 0.33 at 900 K amongst the all HH-x samples. Although the highest *ZT* of HH-1.0 increases by 450% as compared with that of HH-0, the as attained *ZT* value is still lower than that of the *n*-type nanocomposites or the benchmark *ZT* ~ 1 for practical TE materials. For the future work, doping will be implemented in conjunction with the nanostructuring process.

## Acknowledgements

The work at Wuhan University of Technology is supported by the International Science & Technology Cooperation Program of China (Grant No. 2011DFB60150), National Basic Research Program of China (Grant No. 2007CB607501), as well as 111 Project (Grant No. B07040). The work at EMPA is supported by Swisselectric. The work at Clemson University (J.H and T.M.T) is supported by a DOE/EPSCoR Implementation Grant (#DE-FG02-04ER-46139), and the SC EPSCoR cost-sharing program.). W.J.X would like to thank the Marie Curie COFUND fellowship supported by EU FY7 and EMPA.

**Figure captions**

**Figure 1**. XRD patterns of as prepared TiCo$_{0.85}$Fe$_{0.15}$Sb-x%InSb (x=0, 0.7, 1.0, 1.5, and 3.0) composites. In this figure, HH represents the formula 'TiCo$_{0.85}$Fe$_{0.15}$Sb', and x represents the content of InSb.

**Figure 2**. SEM images of HH-0 (a and b) and HH-3 (c and d). In (d), the squared area shows a large population of InSb nano-precipitates.

**Figure 3.** (a) Temperature dependences of the electrical conductivity; (b) InSb content dependences of carrier concentration and carrier mobility; (c) Temperature dependences of the Seebeck coefficient; (d) The room temperature Seebeck coefficient as a function of carrier concentration, with the solid line representing an $\alpha \sim \ln(n^{-1})$ dependence as expected from Equation (1).

**Figure 4.** Temperature dependence of the power factor for HH-x samples.

**Figure 5**. Temperature dependences of (a) thermal conductivity and (b) lattice thermal conductivity. The insets in (a) and (b) highlight the detail from 600 to 900 K. In (b), the blue line (lower left corner) represents the $\kappa_L \propto T^3$ dependence (Debye model), and the red line represents the $\kappa_L \propto 1/T$ dependence (Umklapp processes).

**Figure 6.** Temperature dependence of $C_P$ for HH-0 and HH-0.7, and the blue solid line represents the Dulong-Petit estimation.

**Figure 7.** (a) Temperature dependence of phonon mean free path ($l_{ph}$) for HH-0 and HH-0.7; (b) the detail of $l_{ph}$ from 300 to 900 K, and the blue solid line represents the minimum phonon mean free path (lattice parameter) for TiCoSb materials.

**Figure 8.** The blue solid line represents the theoretical minimum thermal conductivity of TiCoSb calculated from Cahill's model [32]. The lattice thermal conductivity of HH-1.5 is also included for comparison.

**Figure 9.** (a) Temperature dependence of ZT values for HH-x samples; (b) The *ZT* dependence of $\mu/\kappa_L$. The broken line is a guide to the eye.

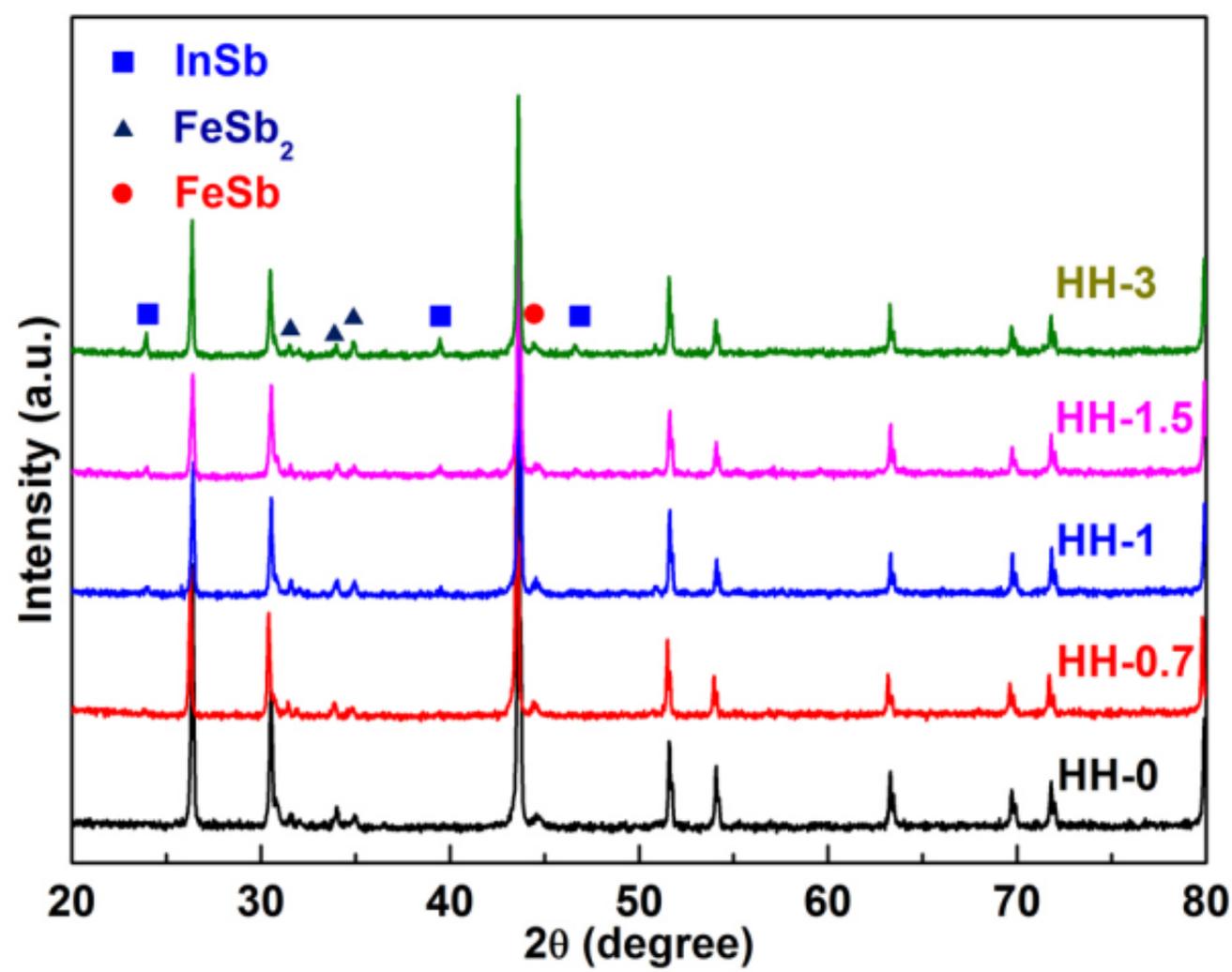

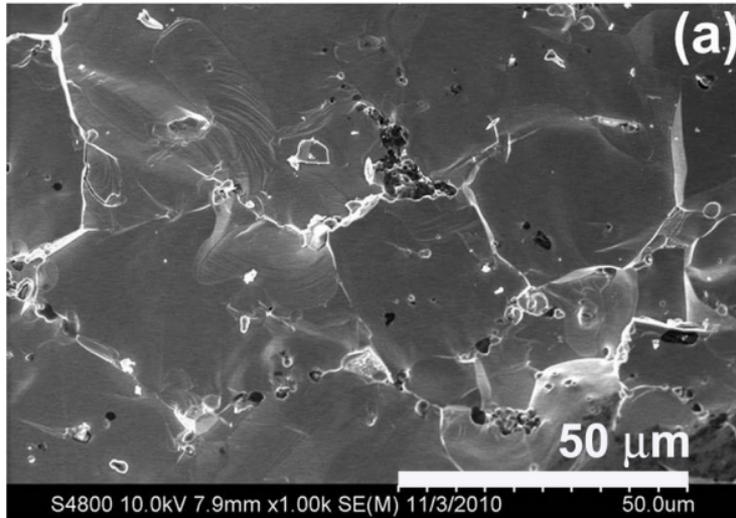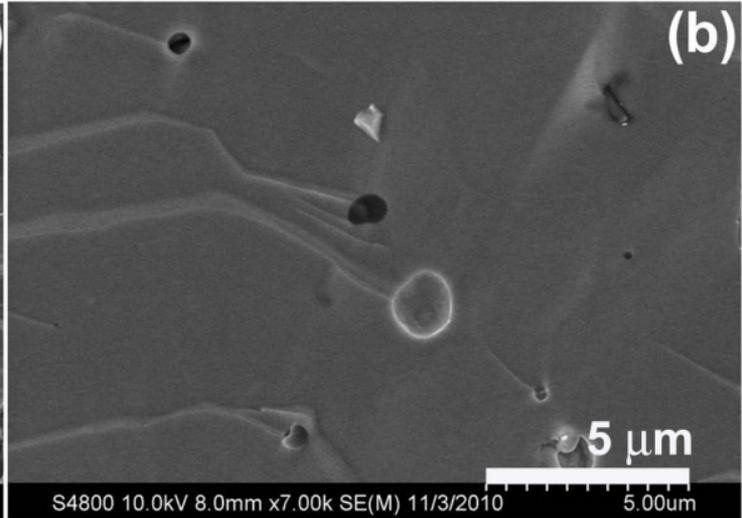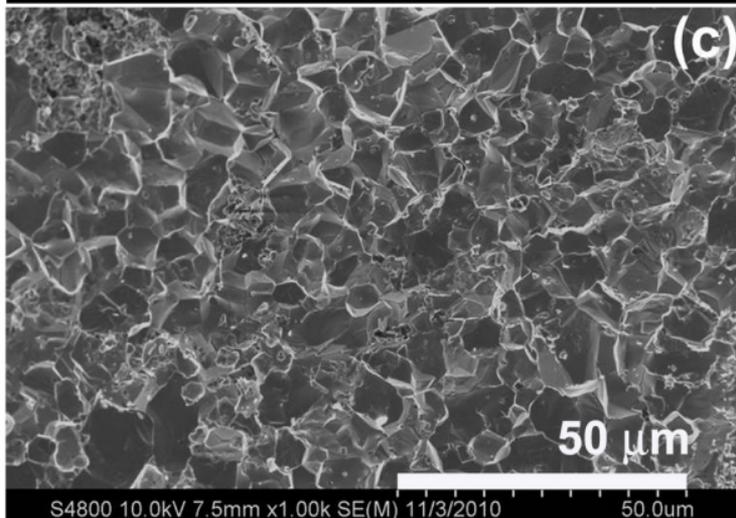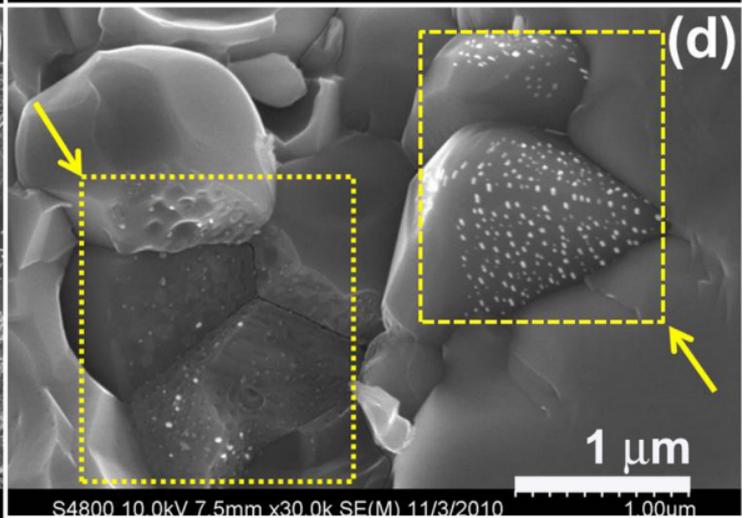

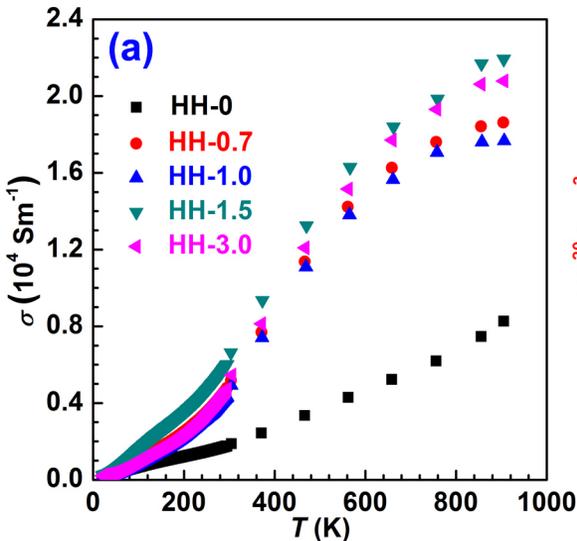
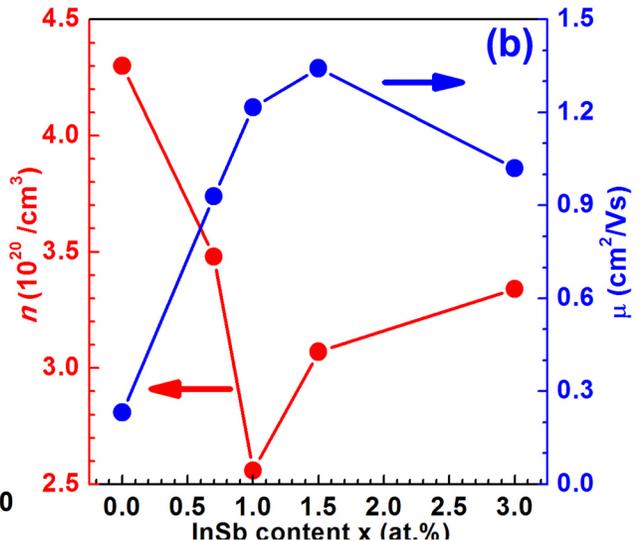
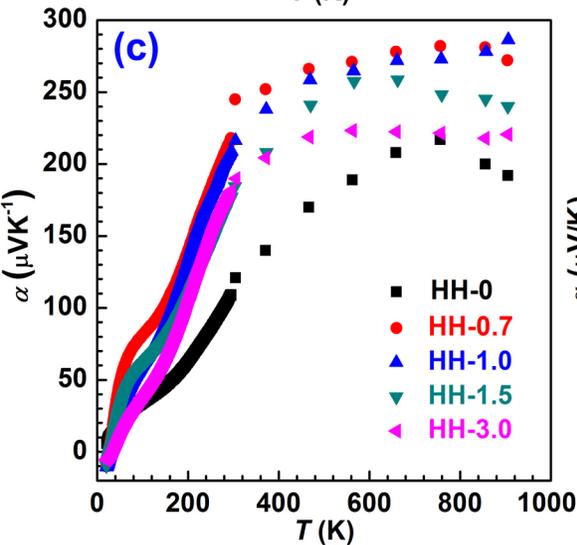
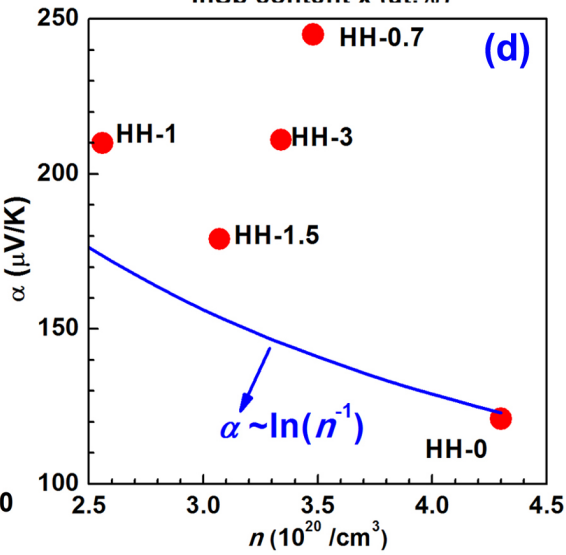

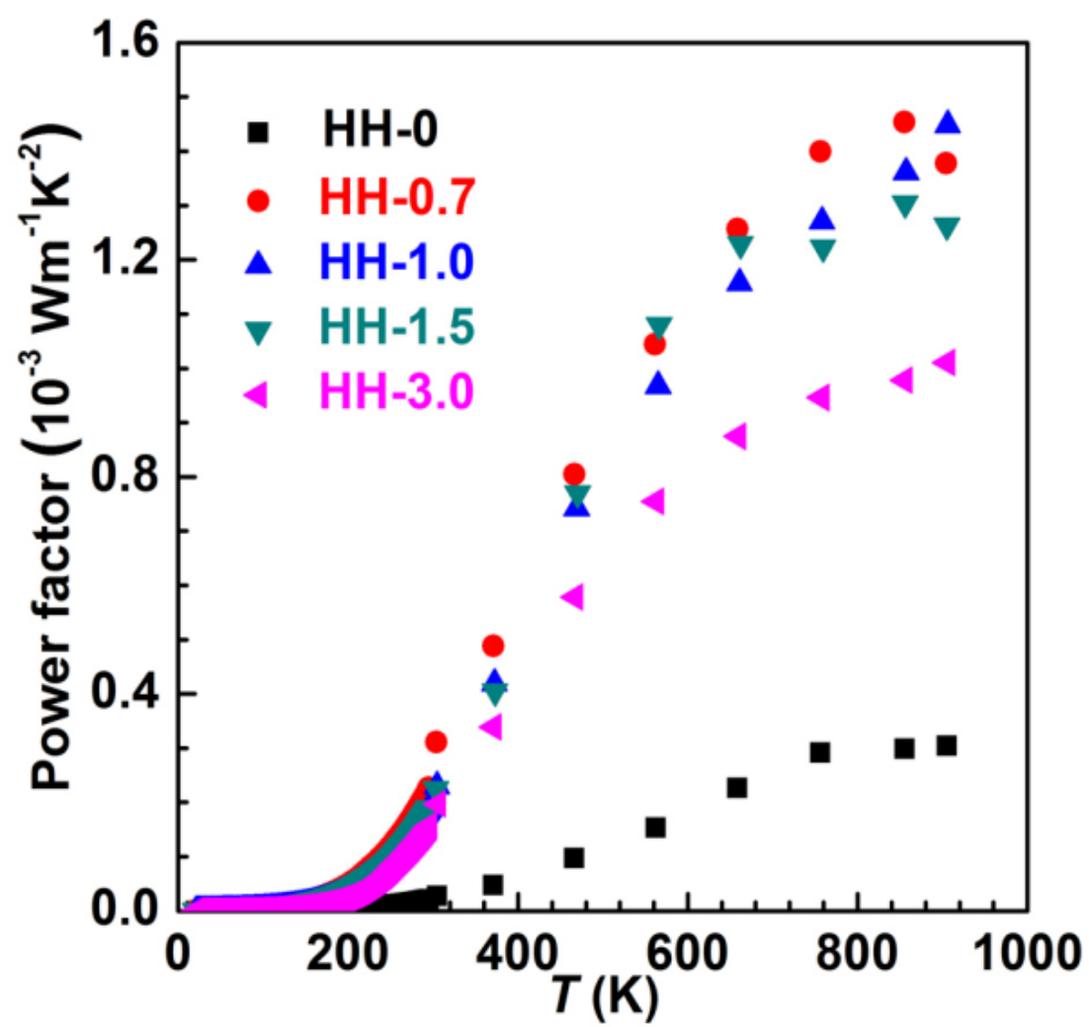

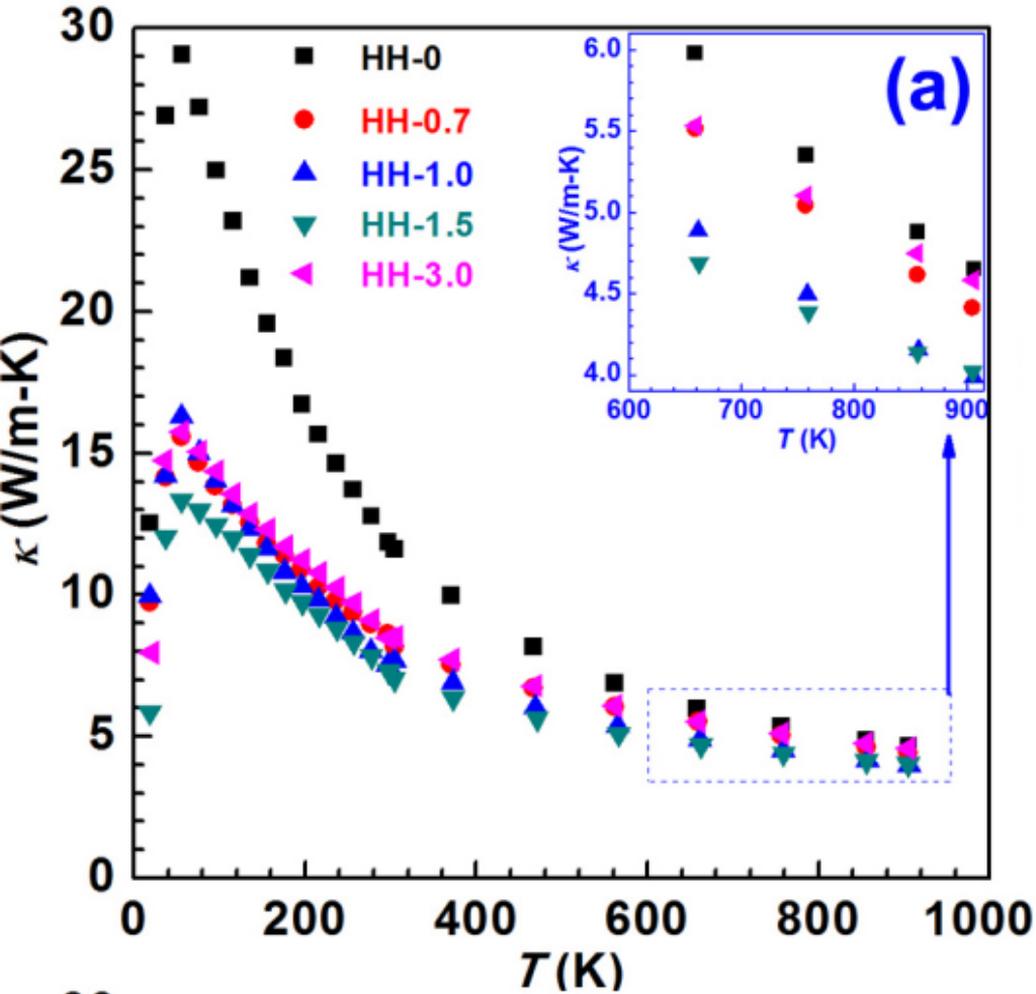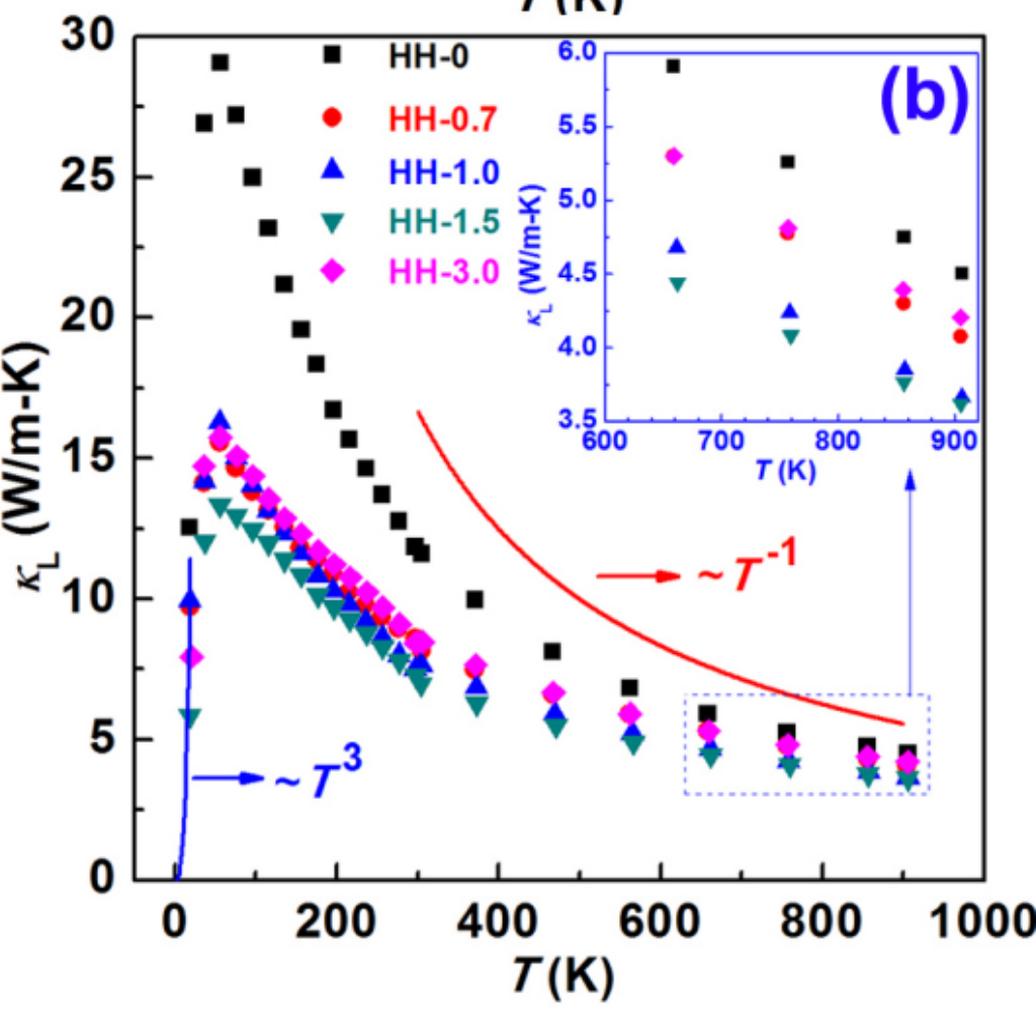

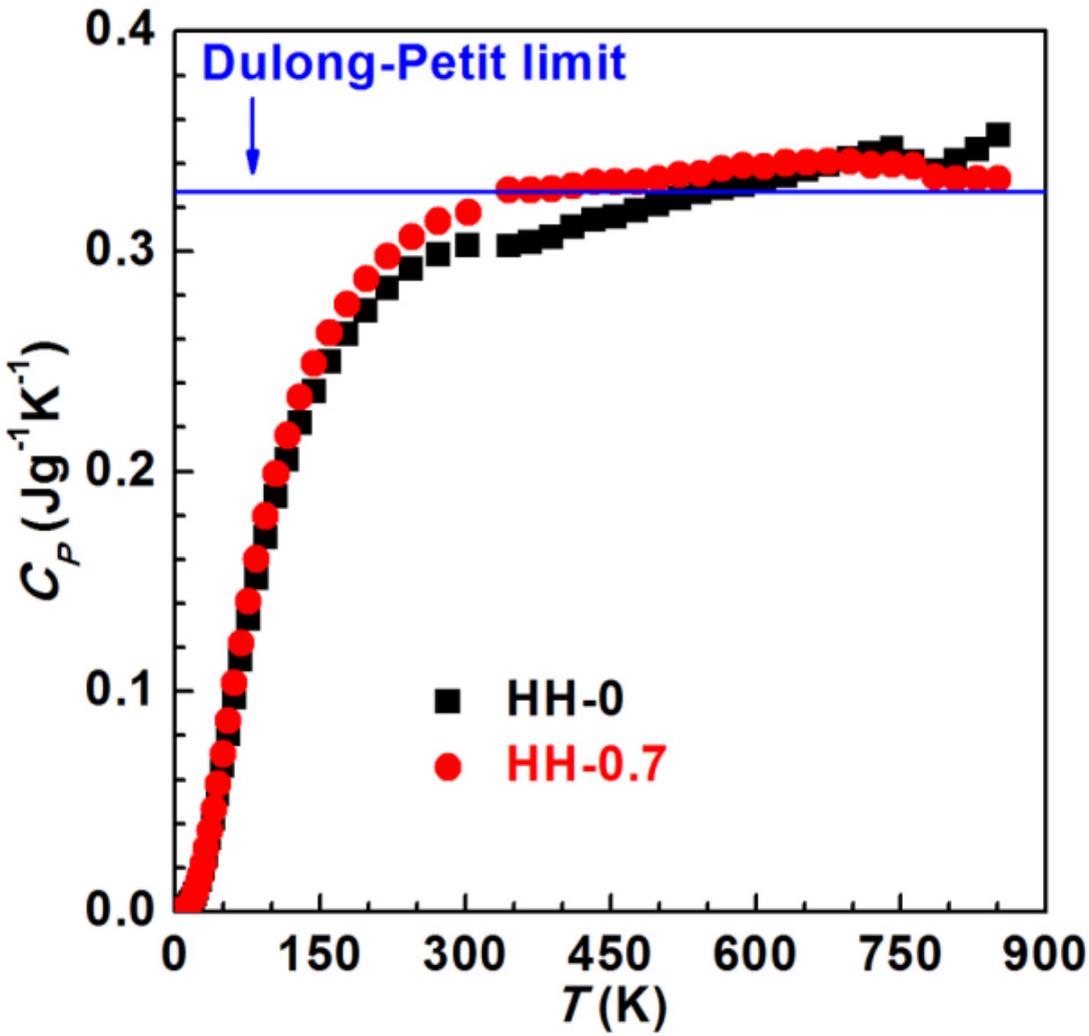

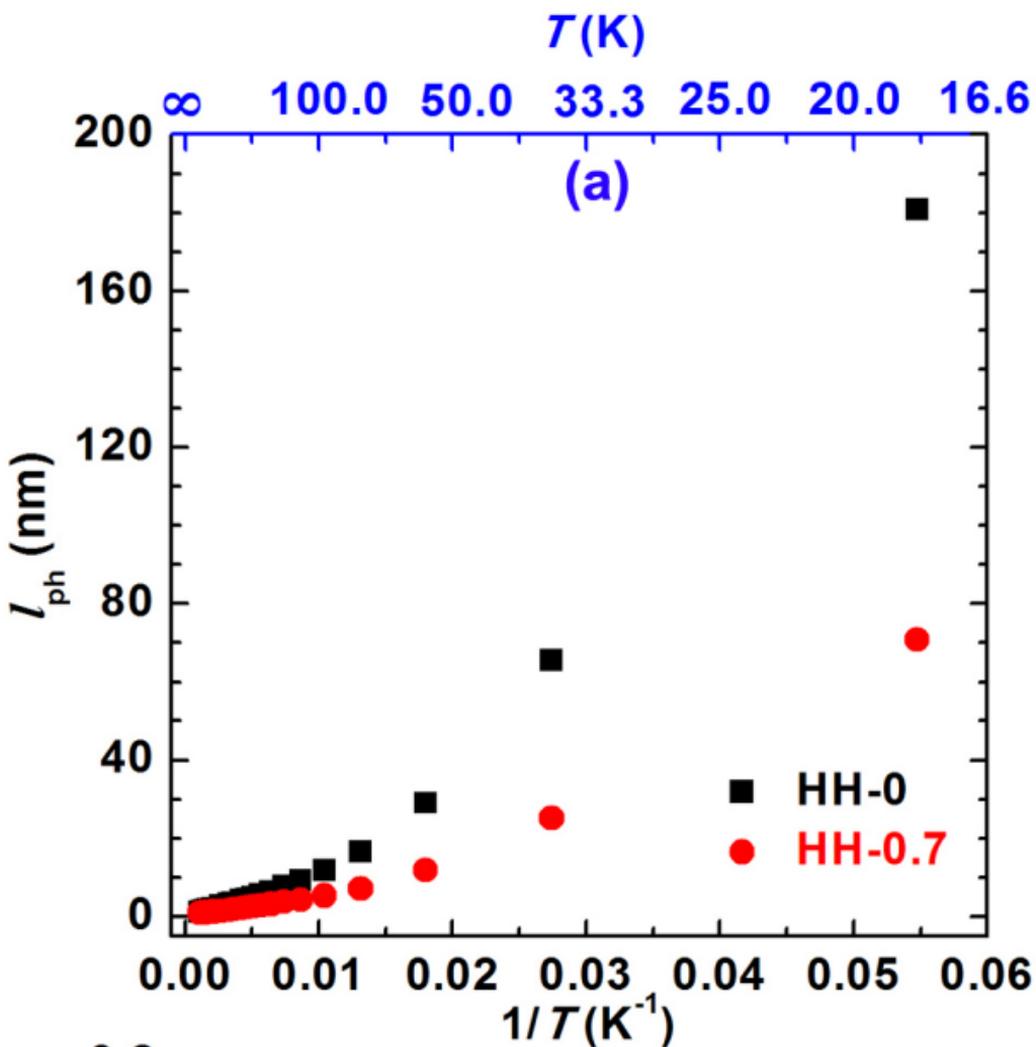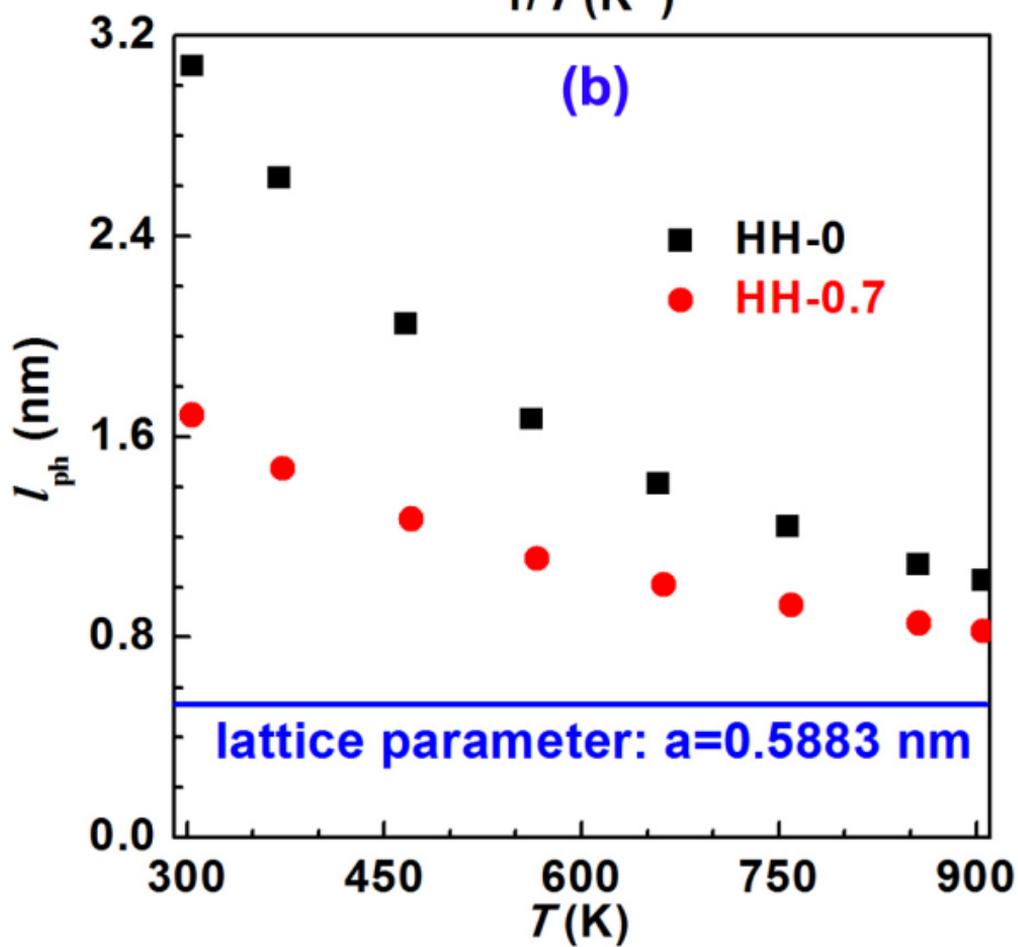

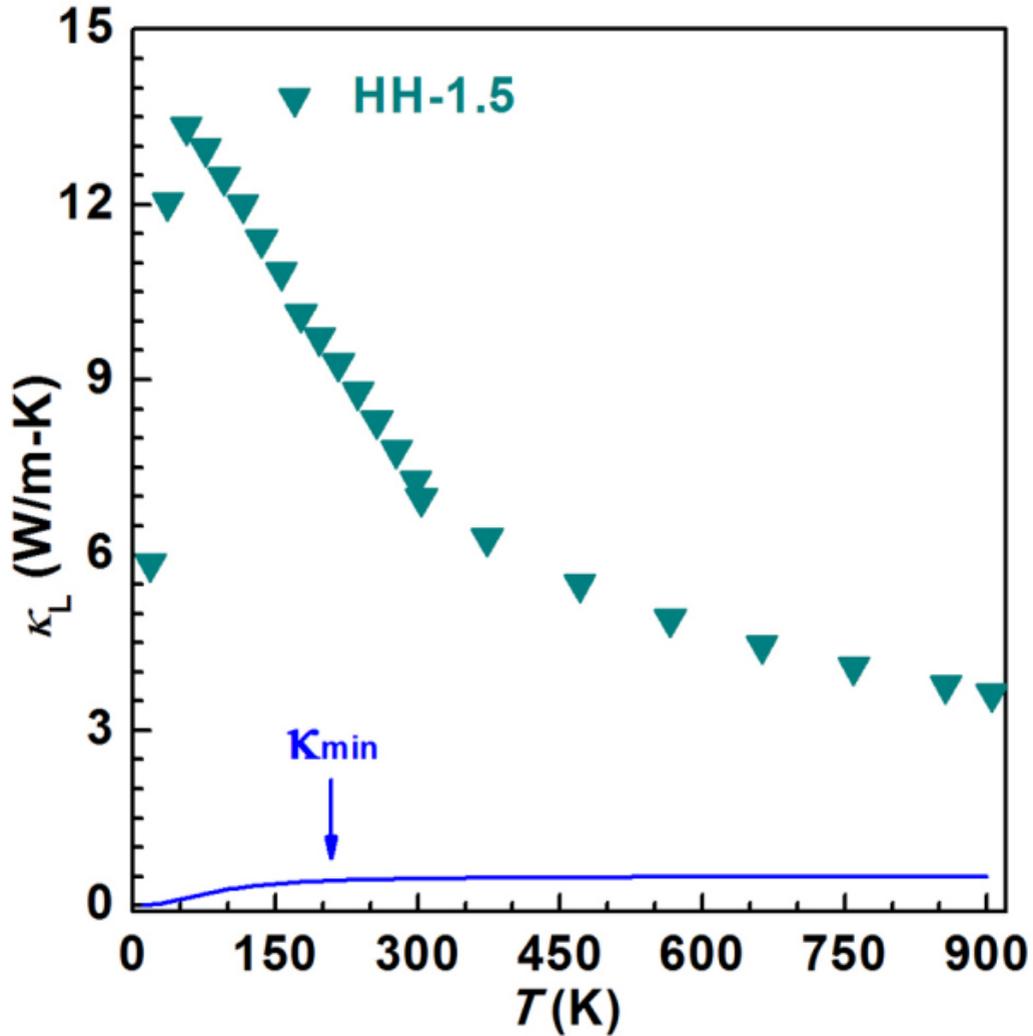

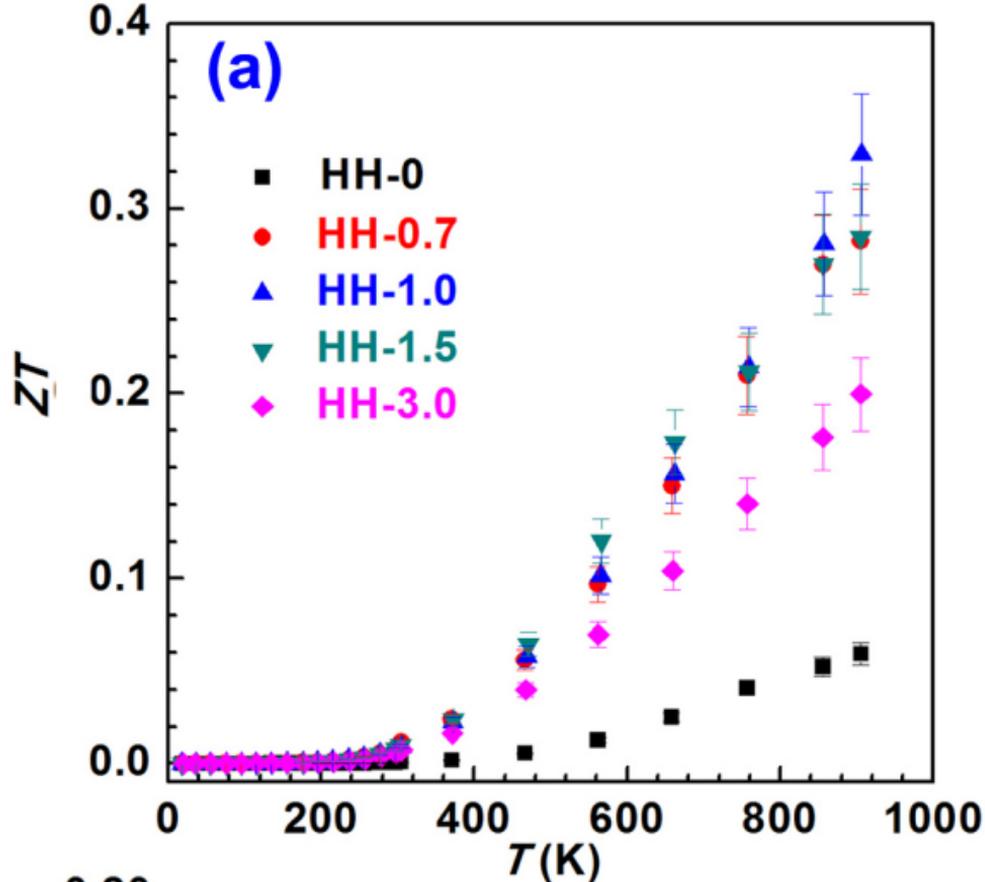
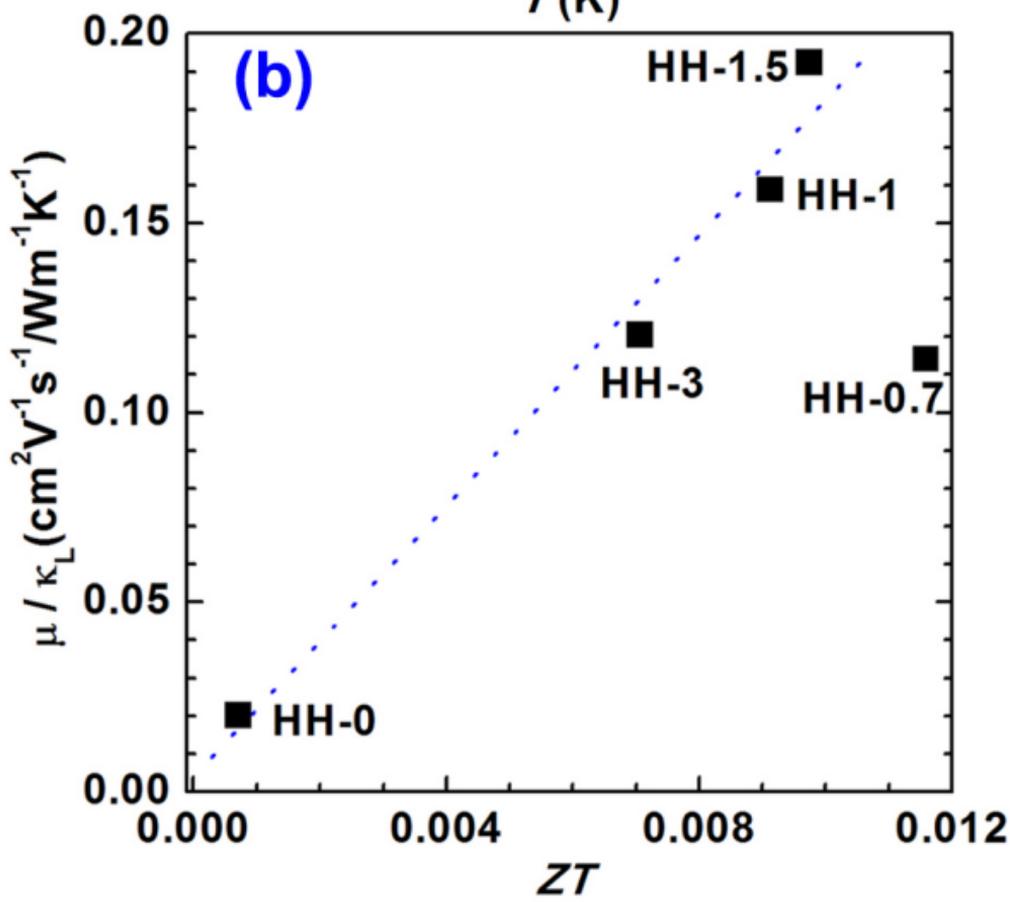

# Significant *ZT* Enhancement in p-type Ti(Co,Fe)Sb-InSb Nanocomposites via a Synergistic '*High Mobility Electron Injection, Energy filtering and Boundary Scattering*' Approach

Wenjie Xie, Yonggao Yan, Song Zhu, Menghan Zhou, Sascha Populoh, Krzysztof Gałązka, S. Joseph Poon, Anke Weidenkaff, Jian He, Xinfeng Tang, and Terry M. Tritt

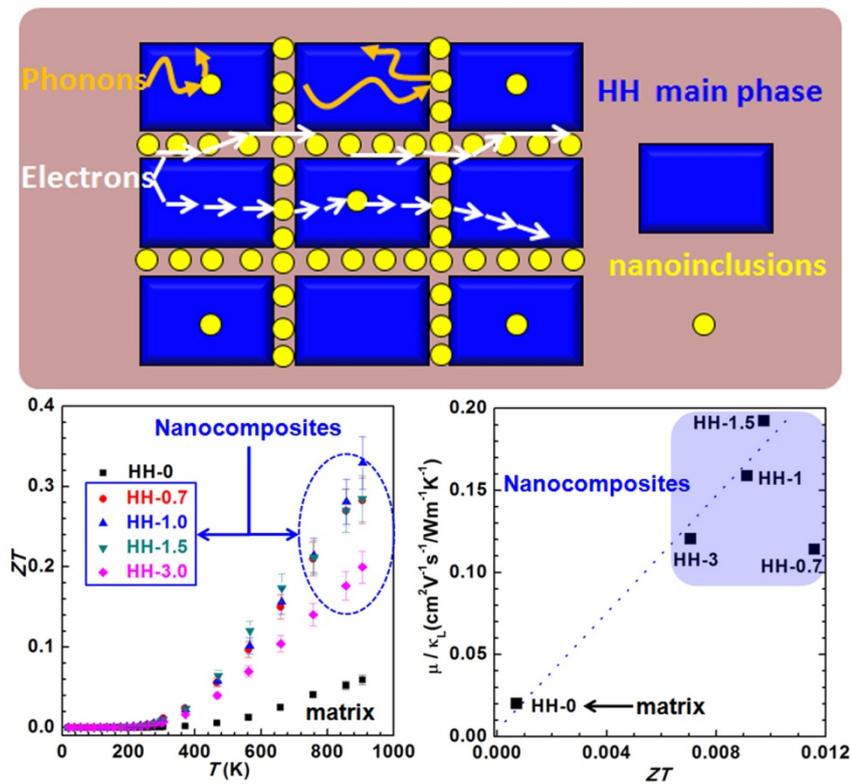

*In-situ* formed InSb nanoinclusions in p-type Ti(Co,Fe)Sb half-Heusler compound can induce combined *high mobility electron injection, low energy electron filtering, and boundary scattering* effects, and lead to a simultaneous improvement of all three individual thermoelectric properties of Ti(Co,Fe)Sb-InSb nanocomposites: enhanced Seebeck coefficient and electrical conductivity as well as reduced lattice thermal conductivity.